\documentclass[conference]{IEEEtran}

\IEEEoverridecommandlockouts
\PassOptionsToPackage{draft}{hyperref}

\usepackage{array}
\newcolumntype{C}[1]{>{\centering\arraybackslash}m{#1}}
\usepackage{makecell}
\usepackage{cite}
\usepackage{hyperref}
\hypersetup{
	colorlinks=false,
	pdfborder={0 0 0},
}
\usepackage[pdftex]{graphicx}
\usepackage{url}
\usepackage{soul}
\usepackage{tabularx}
\newcolumntype{L}{>{\raggedright\arraybackslash}X}
\usepackage{multirow}

\usepackage{siunitx}
\usepackage{tikz} 
\usepackage{pgfplots}
\pgfplotsset{compat=1.12}

\pgfplotsset{compat=newest} 
\usepgfplotslibrary{units}

\usepackage{listings}
\usepackage{fancyvrb}
\usepackage{framed}
\usepackage[listings,skins]{tcolorbox}
\usepackage[skipbelow=\topskip,skipabove=\topskip]{mdframed}
\mdfsetup{roundcorner=0}

\usepackage{textcomp}

\renewcommand\IEEEkeywordsname{Keywords}

\usepackage{xpatch}
\makeatletter
\patchcmd\@makecaption{\\}{.~}{}{\fail}
\makeatletter

\begin{document}
	
	\title{Generating Mock Skeletons for Lightweight Web-Service Testing}
	
	\author{\IEEEauthorblockN{Thilini Bhagya}
		\IEEEauthorblockA{
			\textit{Massey University}\\
			Palmerston North, New Zealand\\
			t.bhagya@massey.ac.nz}
		\and
		\IEEEauthorblockN{Jens Dietrich}
		\IEEEauthorblockA{
			\textit{Victoria University of Wellington}\\
			Wellington, New Zealand\\
			jens.dietrich@vuw.ac.nz}
		\and
		\IEEEauthorblockN{Hans Guesgen}
		\IEEEauthorblockA{
			\textit{Massey University}\\
			Palmerston North, New Zealand\\
			h.w.guesgen@massey.ac.nz}}
	
	\maketitle
	
	\begin{abstract}
		Modern application development allows applications to be composed using lightweight HTTP services. Testing such an application requires the availability of services that the application makes requests to. However,  access to dependent services during testing may be restrained. Simulating the behaviour of such services is, therefore, useful to address their absence and move on application testing. This paper examines the appropriateness of Symbolic Machine Learning algorithms to automatically synthesise HTTP services' mock skeletons from network traffic recordings. These skeletons can then be customised to create mocks that can generate service responses suitable for testing. The mock skeletons have human-readable logic for key aspects of service responses, such as headers and status codes, and are highly accurate. 
	\end{abstract}
	\bigskip
	
	\section{Introduction}
	Service-Oriented Computing (SOC)~\cite{papazoglou2003service} is a popular computing paradigm that supports accelerated, low-cost development of distributed applications in heterogeneous environments. There is a range of Web service technologies that have been used in SOC, starting with early attempts such as the Web Services Description Language (WSDL)~\cite{christensen2001web} and Simple Object Access Protocol (SOAP)~\cite{box1999soap}. More recently, lightweight HTTP-based services,  (i.e., \textit{RESTful services}~\cite{fielding2000architectural}) have become the mainstream.
	
	\medskip 
	When utilising HTTP services, diverse parts of the application interact by sending and responding to HTTP requests, to access and manipulate resources. It is, therefore, easy to implement both clients and servers using a wide range of languages and deploy them on various platforms. In addition, this approach is being increasingly adopted to facilitate the development of product ecosystems around services. Many of the successful services like Google\footnote{\url{https://developers.google.com/apis-explorer/} [accessed Jul. 2 2019]}, Twitter\footnote{\url{https://developer.twitter.com/en/docs.html} [accessed Jul. 2 2019]}, and Flickr\footnote{\url{https://flickr.com/services/api/} [accessed Jul. 2 2019]} currently offer APIs which provide simple access to their resources and services.
	
	\medskip 
	One of the key challenges when developing applications using services is to assure the accurate functioning before deploying. Adequately testing with dependent services is not always possible due to limitations in the observability of service code, lack of control, and the costs of accessing services~\cite{papazoglou2003service}. Therefore, researchers in both industry and academia have started to investigate applicable approaches to test such applications independent of the services which they depend on.
	
	\medskip 
	A common practice is \textit{mocking}, i.e., to \textit{manually} define behavioural responses from scratch based on underlying service syntax and semantics (i.e., mock objects~\cite{freeman2004mock}). A challenge of mocking is that it requires a detailed understanding of the service semantics. Service virtualisation (SV)~\cite{michelsen2012service} tries to address this problem by \textit{automatically} constructing virtual models of services suitable for testing by inferring service semantics from traffic recordings. For example, SV will try to emulate the behaviour of a dependent service by generating anticipated responses using the inferred semantic model. 
	
	\medskip
	SV often draws on artificial intelligence (AI) techniques as inference mechanisms. Amongst them, symbolic machine learning (SML) algorithms have gained popularity due to the provenance of their results, i.e., humans can understand the outcome of the results with relative ease as the system produces a human-readable explanation. This is in stark contrast to many sub-symbolic AI techniques that lack provenance, and are black-box by nature. Example for symbolic AI techniques are representations like decision-trees and logical rules~\cite{michalski2013machine}. While these techniques are well-established, they have gained attention recently within the context of explainable-AI~\cite{holzinger2018machine}.
	
	\medskip 
	The approach proposed here can be seen as a hybrid technique: we propose to use SV to infer some attributes of HTTP service responses, but acknowledge that engineers will often want to fine-tune this. This requires the SV algorithm to produce results that can be understood and customised by engineers. We hypothesise that if the SV is based on inference rules, then engineers used to writing mock tests will find it easy to customise those rules. To emphasise this point, consider the snippet of Java code in Fig.~\ref{fig:code} written using the popular mockito framework\footnote{\url{https://site.mockito.org/} [accessed Jul. 25 2019]}. In the second line, the functionality of a linked list is \textit{stubbed} for the purpose of testing. Interestingly, no actual \texttt{LinkedList} is required at this stage. The process of stubbing basically uses a simple logical rule, expressed using a domain-specific language provided by the mock framework. By using explainable SML techniques, we aim at a solution that provides a sweet spot between highly accurate automation, and customisability. This takes into account that completely automated techniques are unlikely to provide sufficient accuracy to mock complex services. Consider for instance a service providing financial transactions: while it is certainly feasible to infer rules modeling the response codes of accessing account  information, based  on authentication headers, URLs, resource ids and state inferred from transaction history, it is much more complex to model unexpected server behaviour (such as a server returning status code 500) and in general the flakiness associated with distributed and concurrent systems, or the content of data returned by the server (such as inferring the structure of PDF documents synthesised by the server, used for account statements).
	\begin{figure}
		\begin{lstlisting}[deletekeywords={class}] 
		LinkedList mockedList = mock(LinkedList.class);
		when(mockedList.get(0)).thenReturn("first");
		\end{lstlisting}
		\caption{A sample stubbing method call}
		\label{fig:code}
	\end{figure}
	
	\medskip 
	Therefore, the aim of this study is to understand the appropriateness of SML techniques for generating mock skeletons of HTTP services directly from traffic records. We consider three techniques, the decision tree algorithm C4.5 and the rule learners RIPPER and PART. All the experiments have been done within the Waikato Environment for Knowledge Analysis (WEKA) environment~\cite{hall2009weka}, employing network traffic datasets extracted from a few different successful, large-scale HTTP services (i.e., GitHub, Twitter, Google and Slack).
	
	\medskip 
	The rest of this paper is organised as follows: Section~\ref{sec:relatedwork} provides a brief review of related work. Section~\ref{sec:datasets} outlines the datasets used in the experiments. Section~\ref{sec:algo} introduces the decision tree and rule learning algorithms used. Section~\ref{sec:methodology} discusses the procedure that has been adopted. Section~\ref{sec:results} presents the results obtained by classifiers. Section~\ref{sec:validity} reviews some threats to validity. Finally, conclusions are pointed out in Section~\ref{sec:conclusion}.
	
	\section{Related Work} \label{sec:relatedwork}
	
	Service virtualisation (SV) is used in the software industry for addressing dependency constraints in application testing. SV solutions simulate the behaviour of dependent services through synthesising responses mainly by recording and then replaying interaction messages between the application-under-test (AUT) and the live service, assuring the development and test teams have continuous access to realistic test environments. There are multiple vendors providing SV tooling, including  Parasoft~\cite{parasoft}, CA~\cite{ca}, and some open sources projects (e.g., Wiremock~\cite{wiremock}, Hoverfly~\cite{hoverfly}). Most of these support multiple protocols. However, all rely on a priori knowledge of the service structure and message protocol. Possibly responses are manually modified after the recording (i.e., when responses are based on request data). Also, the quality of synthesised responses depends on the availability of traffic recordings for every potential interactive scenario. Plus, the tools operate as black boxes (users cannot understand how a particular service produce responses). Table I shows a general evaluation of these tools. 
	\renewcommand\theadalign{bc}
	\renewcommand\theadfont{\bfseries}
	\renewcommand\theadgape{\Gape[1pt]}
	\renewcommand\cellgape{\Gape[1pt]}
	\renewcommand{\cellalign}{l}
	\begin{table}[!t]
		\caption{Comparison of Service Virtualisation Tools}
		\label{tab:tool:overview}
		\renewcommand{\arraystretch}{1.3}
		\centering
		\begin{tabular}{|l|l|l|l|l|}
			\hline
			\thead{Tool} & \thead{Protocol\\Supported} & \thead{Reasons \\on State} & \thead{Commercial} & \thead{Uses AI} \\	
			\hline
			Parasoft & Most & Yes & Yes & No\\
			\hline
			CA & Most & Yes & Yes & Yes \\
			\hline
			Wiremock & HTTP & Yes & No & No\\
			\hline
			Hoverfly & HTTP & Yes & No & No\\
			\hline    
		\end{tabular}
	\end{table}
	
	\medskip 
	Opaque SV~\cite{du2016opaque,versteeg2017entropy} is a proposal where dependent services are emulated by synthesising responses using a semantic model of a service inferred from recorded interactions. It allows responses to be created automatically, without requiring prior knowledge of the service protocols. The inference is done by means of supervised machine learning techniques (i.e., uses Multiple Sequence Alignment to derive message prototypes from recorded traffic and the Needleman-Wunsch algorithm~\cite{needleman1970general} to match incoming requests against prototypes to generate responses). FancyMock~\cite{eniser2018fancymock} is similar to Opaque SV, but can handle messages with an arbitrary message format. However, all authors ignored the temporal properties of protocols when formulating responses. That is, response generation solely depending on the incoming request and the recorded interaction traces, but not the service state history. Therefore these techniques are only adequate when the target service is stateless or where the testing scenario does not require highly accurate responses. Also, SV results that are generated using all these techniques lack provenance.
	
	\medskip 
	A recent study by Eni{\c{s}}er et al~\cite{enicser2018testing} proposes two different techniques to simulate a service behaviour considering interactions history (state). One of the solutions proposed employs the RIPPER classification algorithm (which we also use in our work), the other one uses neural networks. The classification technique applies one-hot encoding for input data (use up to history size of 10 and incoming request) and applies RIPPER to construct models to predict the response for a given request. This technique performs better in terms of training time but virtual services trained using neural networks produce the more accurate responses. While this work is close to the study presented in this paper, there are some important differences. Their approach is lacking the possibility for reliable prediction of HTTP-based services. The technique is biased towards predicting the response status. It does not incorporate all service features (i.e., headers) when constructing classifiers. The datasets used are not satisfactory as experiments are small and may miss important aspects of real-world, state-of-the-art services.
	
	\medskip 
	Service-oriented applications testing has been extensively examined in the literature. Like, for example,~\cite{bozkurt2010testing,kumar2015empirical,canfora2007service} covers broad surveys on Web service testing. In the context of RESTful Web services, Arcuri~\cite{arcuri2018evomaster} proposes a tool called EvoMaster to automatically collect and exploit white-box information from API specifications and code instrumentation to generate system-level test cases using evolutionary algorithms, Seijas et
	al.~\cite{lamela2013towards} presents a technique to generate tests on a property-based model that depicts an idealised form of RESTful services, Chakrabarti and Rodriquez~\cite{lamela2013towards} also presents a method to generate tests based on a model that represents connectedness of RESTful services, etc.
	
	\medskip 
	AI techniques have been widely used for automating application testing processes. Examples include the work of Vanmali et al.~\cite{singhal2014generation} on how neural networks and decision trees can be used toward implementing test oracles. The classification models learned from input test cases can predict the expected behaviour of AUT (new test cases). The work of Briand et al.~\cite{briand2008using} proposes a methodology to re-engineering test suites based on rules induced from C4.5 tree algorithm that relate input properties to output equivalence classes. These rules can be analysed to determine potential improvements in test suites. Kanewala's and Bieman's~\cite{kanewala2016predicting} presents a technique based on the C4.5 tree algorithm to automatically predict metamorphic relations for automating the testing process without test oracles. AI techniques have also been used to generate statistical test oracles for performance regression testing~\cite{hewson2015performance}.
	
	\medskip 
	Fuzz testing is an increasingly popular testing technique. While first-generation black-box fuzzers generated random input in order to expose defects, modern grey-box and white-box fuzzers use or even infer models of the application under test in order to increase the chances of discovering folds. An example is fuzzers that can infer the grammar of the programming language or data format~\cite{hoschele2016mining}, or use dynamic feedback from test executions (such as coverage). Recently, some of those ideas have been applied to RESTful Web services testing~\cite{atlidakis2019restler}. Fuzzing uses AI to generate the actual tests, whereas our approach uses AI to generate services the (user-written) tests interact with, and we use fuzzing techniques to construct the datasets for evaluation, this will be discussed in Section~\ref{sec:datasets}.
	
	\section{Dataset Acquisition}\label{sec:datasets}
	GHTraffic~\cite{bhagya2018ghtraffic} is a publicly available HTTP dataset, designed for experimenting on various aspects of service-oriented
	computing. The authors extracted the base dataset from GitHub, by reverse-engineering API interactions from an existing repository snapshot and further augmented it with synthetic transactions to include interactions that cannot be recovered from snapshots. All transactions complied with the syntax and semantics of HTTP, and GitHub API specification. It supports a wide range of HTTP features, such as various HTTP methods and status codes. This dataset reflects non-trivial, realistic behavioural patterns of GitHub, plus is in the JSON~\cite{galiegue2013json} format. The small edition of GHTraffic (version 2.0.0)\footnote{\url{https://zenodo.org/record/1034573} [accessed Oct. 3 2019]} was selected for use in this study. 
	
	\medskip 
	In addition to GHTraffic, three HTTP datasets\footnote{\url{https://zenodo.org/record/3378401} [accessed Oct. 3 2019]} were generated by creating random traffic targeting the services offered by Twitter, Google Tasklists, and Slack. In order to form transactions, various operations to create, read, update, and delete (CRUD) service-specific resources were created, and the respective responses were recorded, simulating service interactions by users through applications. The resources, the operations interacted with are tweets (Twitter), messages (Slack), and tasks (Google Tasklists). 
	
	\medskip 
	The actual input generation used fuzzing techniques. In particular, Apache JMeter\footnote{\url{https://jmeter.apache.org/} [accessed Jul. 2 2019]} was used as it has the functionality to fuzz RESTful services (randomly generate various types of API calls by providing different inputs) and recording interactions in a suitable textual format for further processing. The fuzzing was guided by a light-weight semantic service model provided as Swagger spec\footnote{\url{https://github.com/OAI/OpenAPI-Specification} [accessed Jul. 2 2019]}. Swagger (recently renamed as OpenAPI) has emerged as the standard approach for specifying and documenting HTTP APIs, in a way that is both human and machine-readable. The services used to construct dataset all possess Swagger APIs. Therefore, we were able to use Swagger Codegen\footnote{\url{https://github.com/swagger-api/swagger-codegen} [accessed Jul. 2 2019]} to auto-generate JMeter client stubs. This approach allowed us to automate much of the data generation process. 
		
	\medskip 
	Further details of these four different datasets used in this study are summarised in Table~\ref{tab:datasets:overview} showing their request types, response codes, and transaction counts.
	
	\renewcommand\theadalign{bc}
	\renewcommand\theadfont{\bfseries}
	\renewcommand\theadgape{\Gape[1pt]}
	\renewcommand\cellgape{\Gape[1pt]}
	\renewcommand{\cellalign}{l}
	\begin{table}[!t]
		\caption{Overview of HTTP Datasets}
		\label{tab:datasets:overview}
		\renewcommand{\arraystretch}{1.3}
		\centering
		\begin{tabular}{|l|l|l|l|l|}
			\hline
			\thead{Dataset} & \thead{HTTP Method} & \thead{Response Code} & \thead{Count}\\  
			\hline
			GHTraffic & \makecell{GET, HEAD, POST,\\PATCH, PUT, DELETE} & \makecell{200, 201, 204, \\400, 401, 404, \\422, 500} & 32,216\\
			\hline
			Twitter & GET, POST & 200, 404 &26,053 \\
			\hline
			\makecell{Google\\Tasklists} & \makecell{GET, POST, PATCH, \\DELETE} & \makecell{200, 204, 404, \\503} & 4,702\\
			\hline
			Slack & POST & 200 & 17,422\\
			\hline    
		\end{tabular}
	\end{table}
	
	\section{Decision Tree and Rule Induction Algorithms}\label{sec:algo}
	
	Decision tree algorithms generate classifiers in the form of a tree where each node represents a feature (input attribute), each branch represents a possible value that an input attribute can hold, and each leaf represents a value of the target attribute that is to be predicted. The algorithms for building decision trees have been developed and refined over many years, starting with ID3 (Iterative Dichotomizer 3)~\cite{quinlan1986induction} which employs top-down, greedy search through the space of possible branches with no backtracking. ID3 uses entropy (a measure of uncertainty in a set of instances) and information gain (measure of how much information an attribute gives about the class) to construct a decision tree. The C4.5 algorithm~\cite{quinlan2014c4} is an improved version of ID3. The extra specialities of C4.5 are accounting for missing values, decision tree pruning (solves the over-fitting problem by using a bottom-up technique), allowing both continuous, and discrete features, etc. In addition to aforementioned, there are other decision tree algorithms available, for instance, CART~\cite{breiman1984classification}, ADTree~\cite{freund1999alternating}, and Random Forest~\cite{ho1995random}.
	
	\medskip 
	Rule induction algorithms generate classification models as an ordered set of \textit{if-then} rules called decision lists. PART~\cite{frank1998generating} is a separate-and-conquer rule learner that builds a partial C4.5 decision tree in each iteration and makes the leaf with maximum coverage into a rule. The RIPPER~\cite{cohen1995fast} (Repeated Incremental Pruning to Produce Error Reduction) algorithm uses incremental reduced-error pruning for constructing decision lists. It produces a set of rules by repeatedly adding rules to an empty set until all positive examples are covered. Rules are formed by greedily adding conditions to the antecedent until no negative examples are covered. After constructing a rule set, an optimisation step is performed to reduce its size and enhance its fit to the training data. Besides, several other rule learners such as OneR~\cite{holte1993very}, DecisionTable~\cite{Kohavi1995}, and CN2~\cite{clark1991rule} exist. 
	
	\medskip 
	For our experiments, we needed algorithms that can handle both nominal and numeric predictions, as is required when considering the typical nature of HTTP network traffic datasets (selected datasets mostly contain nominal and numerical values). We also wanted to make sure that we use well-established and effective algorithms for the experiments. These led to the choice of C4.5 decision tree algorithm, and two rule-based algorithms, RIPPER and PART. All three algorithms are widely used to support research in many areas of computer science~\cite{kotsiantis2007supervised}. Apart from being recognised as the most established, and effective decision tree and rule-induction algorithms in use today, selections are made because of their proven ability to deal with nominal and numeric attributes for model building.
	
	\medskip 
	Weka (Waikato Environment for Knowledge Analysis)~\cite{witten2016data} is a popular machine learning and data mining workbench which contains numerous inbuilt algorithms for classification and prediction, accompanying with techniques for pre-processing and post-processing of data. Weka also has a general API to embed other libraries. In this study, experiments are conducted in the WEKA environment by utilising the J48 decision tree classification algorithm (Java implementation of the C4.5 in the Weka), JRip (Weka's implementation of the RIPPER), and PART for constructing classification models. 
	
	\section{Methodology}\label{sec:methodology}
	
\begin{figure}[htbp]
		\centerline{\includegraphics[width=\columnwidth]{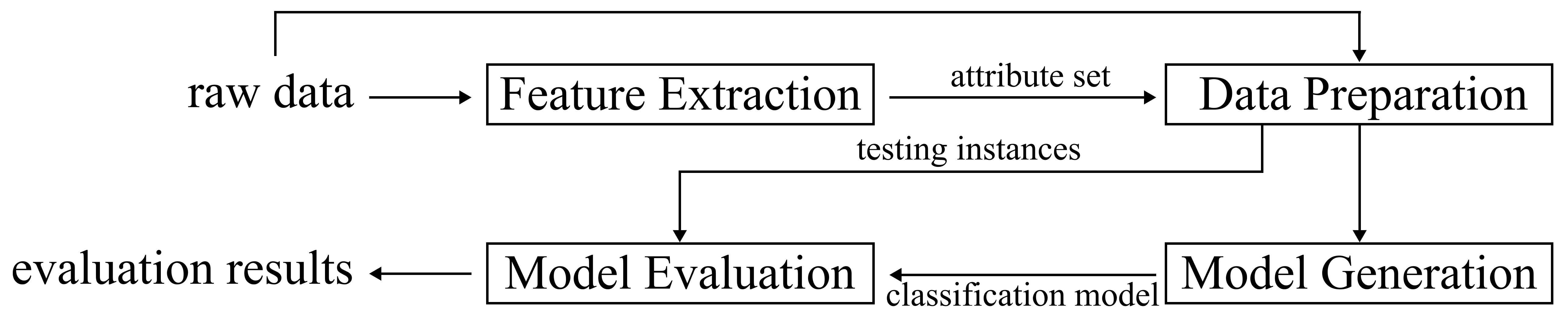}}
		\caption{The processing pipeline}
		\label{fig:processingpipeline}
	\end{figure}

	Fig.~\ref{fig:processingpipeline} provides an overview of the methodology used. In the \textit{Feature Extraction} phase, structural characteristics of HTTP request/response messages will be obtained from the data to collect the attributes that can be used for inference. During the \textit{Data Preparation} phase, the attributes will be filtered and processed. In the \textit{Model Generation} phase, classification models will be built from training data. This will be accomplished by using the respective Weka classifiers (C4.5, PART, and RIPPER). In the \textit{Model Evaluation} phase, the predictive ability of the models generated is assessed.
	
	\subsection{Feature Extraction}
	
\begin{figure}[!t]
		\centerline{\includegraphics[width=\columnwidth]{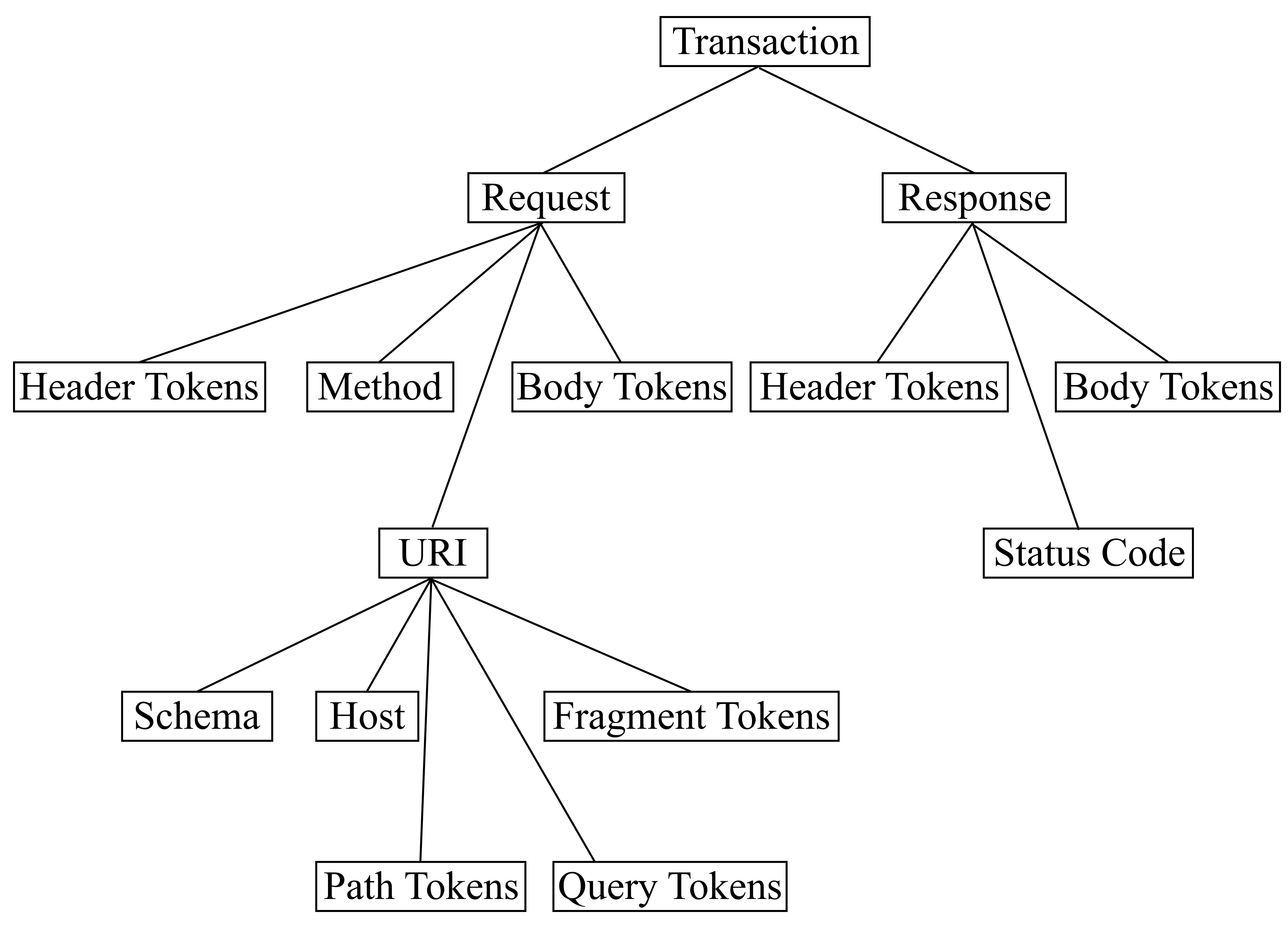}}
		\caption{The feature tree}
		\label{fig:featuretree}
	\end{figure}

	Through interpreting the structural properties of HTTP messages, features were extracted to be used as attributes by the various algorithms. A high-level overview of the features used to capture the message structure of HTTP is shown in Fig.~\ref{fig:featuretree} (features are organised in a simple hierarchy, the \textit{feature tree}). In the following section, we briefly describe the approach used to extract features. 
	
\begin{figure}[!t]
		\centerline{\includegraphics[width=0.81\columnwidth]{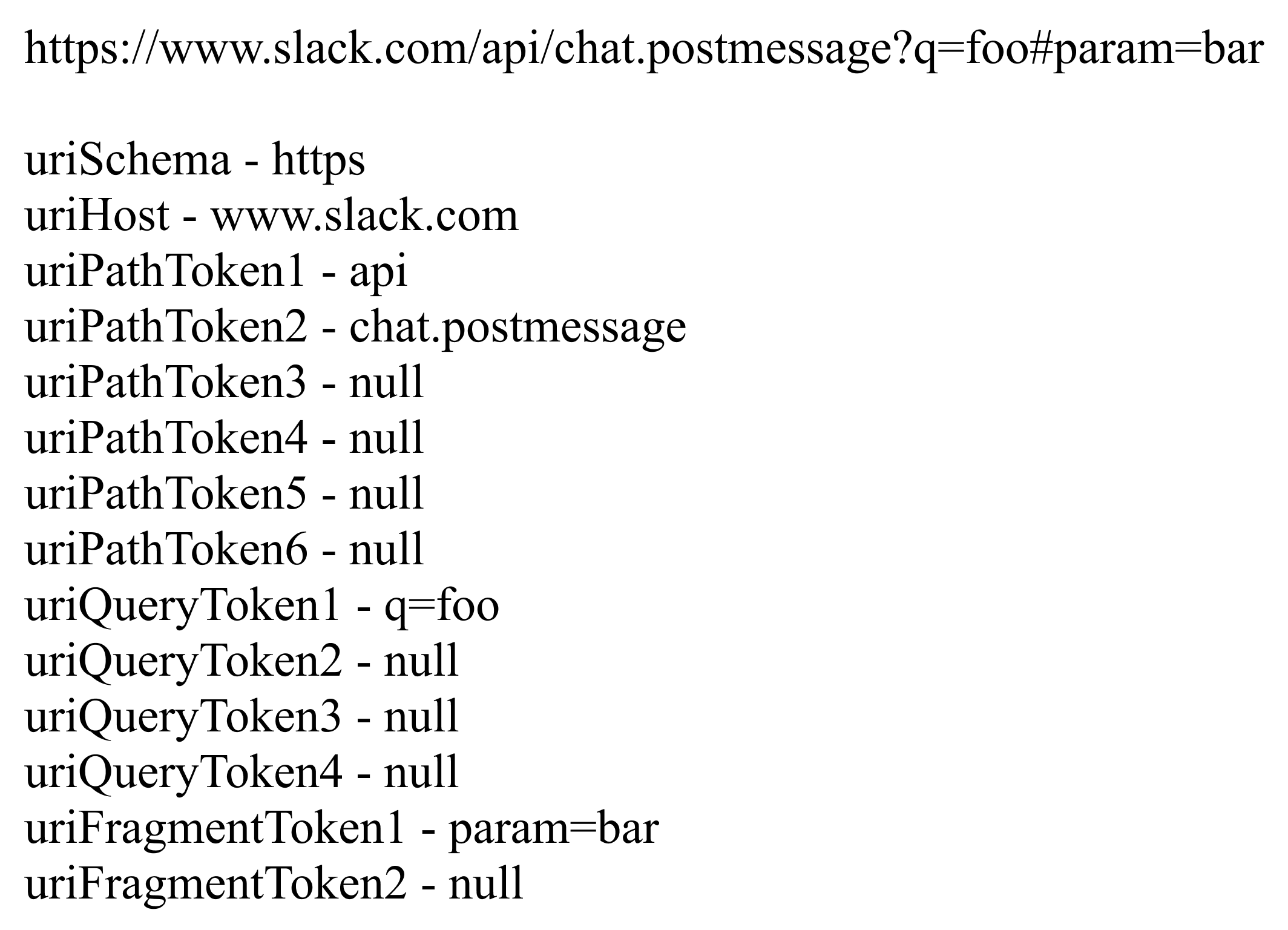}}
		\caption{Attribute and value derivation from a sample URI}
		\label{fig:uri}
	\end{figure}

	\medskip 
	The first category of attributes is general characteristics each HTTP transaction has, in particular, \textit{method} and \textit{statusCode}. Next, we extracted attributes from the request URIs. The URI has a canonical structure ~\cite{berners1998uniform}, consisting of schema, host, path, query, and fragment. Elements like schema and host can be directly used as attributes. The path segment can be tokenised using standard delimiters, (/), and each token can be used as an attribute with a name comprised of \textit{uriPathToken} followed by the position index of the token in the path. If no value is presented for the respective token, the attribute value is set to \textit{null}, otherwise the token value is used. We performed tokenisation on query and fragment components of the HTTP request URI to use as attributes. Fig.~\ref{fig:uri} illustrates this process using an example. 
	
	\medskip 
	It is also possible to obtain extra attributes based on the features of request and response bodies. \textit{hasPayload} and \textit{hasValidPayload} are boolean attributes introduced to represent that requests do have a body and the body is a properly encoded value according to its content type. Although HTTP services can use arbitrary content types, we only support JSON as this is by far the most widely used format for data exchange via HTTP services. Basic tokenisation is done by parsing request and response bodies separately extracting two different sets of all possible service-specific keys given the set of HTTP transactions (usually the body is a JSON object). These keys result in an extensive set of attributes prefixed by \textit{requestjson} or \textit{responsejson} with the key's name. 
	
	\medskip 
	Furthermore, we extracted attributes from HTTP headers, both standard HTTP headers and API-specific headers (i.e., usually identifiable by header names starting with \textit{x-}) in both the request and the response to derive attributes. The process is analogous to the approach described beforehand and applies the same logic with a stipulation for returning \textit{no-exist} if the attribute does not exist in a transaction. In particular, an inferred  \textit{hasAuthorisationToken} attribute is extracted when headers bearing authorisation information are encountered.
	
	\medskip 
	Finally, we generated attributes to represent the features of predecessors on each HTTP transaction (transaction that is just before and set of transactions preceding, over a resource). A boolean attribute \textit{hasImmediatePreviousTransaction} is added to indicate whether or not there was another transaction immediately before, interacting with the same resource. In relation, further attributes are introduced as similar to the approach described earlier. Concerning all predecessors, a set of boolean attributes is obtained to indicate the state of predecessors. Additional boolean attributes are added to identify whether a CRUD operation was performed on a resource, this is based on the use of certain HTTP methods, and/or the presence of certain naming patterns in URI tokens (e.g., Slack uses \textit{postMessage}, \textit{update}, \textit{delete}, Twitter uses \textit{update}, \textit{destroy}, \textit{show}).
	
	\subsection{Data Preparation}
	
	The raw data extracted from recorded HTTP transactions was converted into ARFF (Attribute-Relation File Format) format to use in Weka classifiers. Data type conversion was used as the classification algorithms require feature attribute values to be nominal or binary (numeric or string attributes can have a de facto infinite domain). All numeric values were converted into a set of nominal attributes by applying the \textit{NumericToNominal} filter of Weka. The filter simply takes all numeric values and adds them to the list of nominal values of that attribute (e.g., \textit{statusCode} attribute has a predefined finite set of all possible values after applying the filter). The similar approach is applied to string attributes utilising \textit{StringToNominal} filter.
	
	\medskip 
	The aim of the study is to generate HTTP response skeletons with multiple attributes (e.g. status code, response headers, body). From the machine-learning point of view, this is therefore a multi-class or multi-target learning problem. Unfortunately, Weka currently does not support multivariate predictions. The only option is to train separate models for each target feature. Hence, separate models are trained, one for each target by removing irrelevant target attributes.
	
	\medskip 
	Further classification algorithms have confined the use of classifiers to non-unary targets (the target attribute must have at least two values). Accordingly, all target attributes which have only one distinct value are also ignored as they have no discriminative value. On the other hand, target attributes holding a fairly large set of distinct values can also be  excluded from learning as they are non-optimal for predictions. An example for a feature with only one value is \textit{host} (assuming that a service may always use the same host), an example for a feature with too many values is a high-precision \textit{timestamp} (assuming that each transaction has a unique value). Using such unnecessary features decreases training speed, requires a larger amount of memory, lower model interpretability, and most importantly, can result in overfitting. In all these cases the \textit{Remove} filter of Weka is used to exclude attributes before data is passed to classifiers.
	
	\medskip 
	A summary of the number of features associated with each input dataset after attribute removal is listed in Table \ref{tab:overview:instances}.
	
	\begin{table}[!t]
		\caption{Overview of Input Datasets}
		\label{tab:overview:instances}
		\renewcommand{\arraystretch}{1.3}
		\centering	
		\begin{tabular}{|l|l|l|}
			\hline
			\multicolumn{1}{|c|}{\bfseries Dataset} & \multicolumn{1}{c|}{\bfseries Input Attributes} & \multicolumn{1}{c|}{\bfseries Targets} \\ \hline
			GHTraffic        &    43        &  49       \\ \hline
			Twitter          &      38      &    65     \\ \hline
			Google Tasklists &    42        &  17       \\ \hline
			Slack            &    42        &   8      \\ \hline
		\end{tabular}
	\end{table}
	
	\subsection{Model Generation and Evaluation}
	
	Model generation was performed using Weka 3.5 with the default configuration. The chosen classification algorithms were applied to train multiple models to predict different attributes associated with response properties. 10-fold cross-validation is applied for each classifier on each dataset. The accuracy, precision, and recall were calculated and recorded. Further, the size of the tree (number of nodes it contains) or the number of rules produced by the classifier were measured to assess the comprehensibility (as the number of nodes in a tree is roughly equivalent to the size of the corresponding rule set, this can lead to a reasonable comparison). In order to quantify the overall performance of each algorithm,  mean and the standard deviation were computed from all observed results of the collection of single targeted models. 
	
	\section{Evaluation Results and Discussion} \label{sec:results}
	
	The plot in Fig.~\ref{plot:measures:accuracy} shows the accuracy of each target attribute in Google Tasklists on C4.5, RIPPER, and PART. Table \ref{tab:measures:accuracy}, \ref{tab:measures:precision}, \ref{tab:measures:recall}, and \ref{tab:measures:size} present overall performance measurements (mean and standard deviation) based on different techniques applied, over all four datasets.
	\begin{figure}[!t]
		\centering
		\footnotesize
		\begin{tikzpicture}
		\begin{axis}[	
		ybar,
		grid=major, 
		grid style={dashed,gray!30}, 
		scaled y ticks = true,
		ylabel={\bfseries Accuracy},
		xlabel={\bfseries Response Feature},
		width=0.5*\textwidth,
		height=6cm,
		bar width=2pt,
		symbolic x coords={responseheader:Accept-Ranges,responseheader:Cache-Control,responseheader:Content-Length,responseheader:Content-Type,responseheader:Pragma,responseheader:Transfer-Encoding,responseheader:Vary,responseheader:X-Content-Type-Options,responseheader:X-Frame-Options,responseheader:X-XSS-Protection,statusCode,responsejson:error.code,responsejson:error.errors.domain,responsejson:error.errors.reason,responsejson:error.errors.message,responsejson:error.message,responsejson:kind	
		},
		xtick=data,
		ymin=0.9750,
		enlarge x limits=0.0250,
		tick label style={font=\footnotesize},
		x tick label style={rotate=90,anchor=east,font=\footnotesize},	
		legend style={draw=none,at={(0.625,-1)},anchor=north west,legend columns=-1,font=\tiny}
		]
		\addplot
		coordinates {	
			(responseheader:Accept-Ranges,0.9987)
			(responseheader:Cache-Control,0.9974)
			(responseheader:Content-Length,0.9798)
			(responseheader:Content-Type,0.9998)
			(responseheader:Pragma,0.9983)
			(responseheader:Transfer-Encoding,0.9987)
			(responseheader:Vary,0.9987)
			(responseheader:X-Content-Type-Options,0.9998)
			(responseheader:X-Frame-Options,0.9998)
			(responseheader:X-XSS-Protection,0.9998)
			(statusCode,0.9966)
			(responsejson:error.code,0.9970)
			(responsejson:error.errors.domain,0.9979)
			(responsejson:error.errors.reason,0.9970)
			(responsejson:error.errors.message,0.9970)
			(responsejson:error.message,0.9970)
			(responsejson:kind,0.9977)};
		\addlegendentry{C4.5}
		
		\addplot
		coordinates {
			(responseheader:Accept-Ranges,0.9983)
			(responseheader:Cache-Control,0.9972)
			(responseheader:Content-Length,0.9796)
			(responseheader:Content-Type,0.9998)
			(responseheader:Pragma,0.9981)
			(responseheader:Transfer-Encoding,0.9983)
			(responseheader:Vary,0.9979)
			(responseheader:X-Content-Type-Options,0.9998)
			(responseheader:X-Frame-Options,0.9998)
			(responseheader:X-XSS-Protection,0.9998)
			(statusCode,0.9966)
			(responsejson:error.code,0.9968)
			(responsejson:error.errors.domain,0.9977)
			(responsejson:error.errors.reason,0.9970)
			(responsejson:error.errors.message,0.9968)
			(responsejson:error.message,0.9968)
			(responsejson:kind,0.9977)};
		\addlegendentry{RIPPER}
		
		\addplot
		coordinates {
			(responseheader:Accept-Ranges,0.9987)
			(responseheader:Cache-Control,0.9972)
			(responseheader:Content-Length,0.9798)
			(responseheader:Content-Type,0.9998)
			(responseheader:Pragma,0.9981)
			(responseheader:Transfer-Encoding,0.9987)
			(responseheader:Vary,0.9987)
			(responseheader:X-Content-Type-Options,0.9998)
			(responseheader:X-Frame-Options,0.9998)
			(responseheader:X-XSS-Protection,0.9998)
			(statusCode,0.9966)
			(responsejson:error.code,0.9970)
			(responsejson:error.errors.domain,0.9979)
			(responsejson:error.errors.reason,0.9970)
			(responsejson:error.errors.message,0.9970)
			(responsejson:error.message,0.9970)
			(responsejson:kind,0.9977)};
		\addlegendentry{PART}
		\end{axis}
		\end{tikzpicture}
		\caption{The accuracy per response feature (target) in Google Tasklists. To measure the overall performance on each algorithm in predicting response properties, mean and standard deviation were calculated. See Table~\ref{tab:measures:accuracy}.}
		\label{plot:measures:accuracy}
	\end{figure}
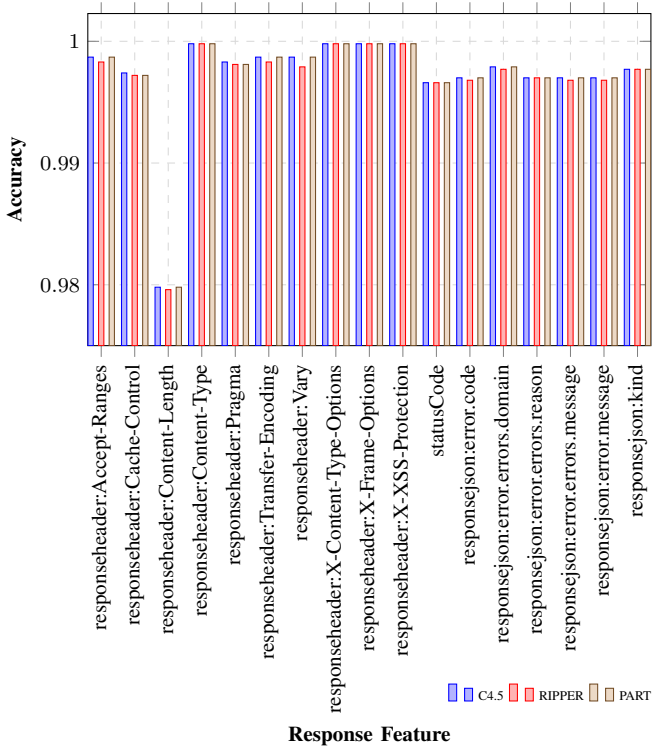
	\begin{table}[!t]
		\caption{Classification Accuracy}
		\label{tab:measures:accuracy}
		\renewcommand{\arraystretch}{1.3}
		\centering
		\begin{tabular}{l|l|l|l|}
			\cline{2-4}
			& \multicolumn{3}{c|}{\bfseries Technique} \\ \hline
			\multicolumn{1}{|c|}{\bfseries Dataset}          & \multicolumn{1}{c|}{\itshape\bfseries C4.5} & \multicolumn{1}{c|}{\itshape\bfseries RIPPER}                      & \multicolumn{1}{c|}{\itshape\bfseries PART}  \\ \hline
			\multicolumn{1}{|l|}{GHTraffic}        &  0.9837$\pm${0.0167}  &  0.9837$\pm${0.0167}       &  0.9834$\pm${0.0170}    \\ \hline
			\multicolumn{1}{|l|}{Twitter}          &  0.9993$\pm${0.0022}  &  0.9993$\pm${0.0018}       &  0.9993$\pm${0.0018}    \\ \hline
			\multicolumn{1}{|l|}{Google}           &  0.9971$\pm${0.0046}  &  0.9969$\pm${0.0046}       &  0.9971$\pm${0.0046}    \\ \hline
			\multicolumn{1}{|l|}{Slack}            &  0.9541$\pm${0.1289}  &  0.9541$\pm${0.1289}       &  0.9541$\pm${0.1289}    \\ \hline
		\end{tabular}
	\end{table}
	\begin{table}[htbp]
		\caption{Classification Precision}
		\label{tab:measures:precision}
		\renewcommand{\arraystretch}{1.3}
		\centering
		\begin{tabular}{l|l|l|l|}
			\cline{2-4}
			& \multicolumn{3}{c|}{\bfseries Technique} \\ \hline
			\multicolumn{1}{|c|}{\bfseries Dataset}          & \multicolumn{1}{c|}{\itshape\bfseries C4.5} & \multicolumn{1}{c|}{\itshape\bfseries RIPPER}                      & \multicolumn{1}{c|}{\itshape\bfseries PART}  \\ \hline
			\multicolumn{1}{|l|}{GHTraffic}        &  0.9687$\pm${0.0332}  &  0.9693$\pm${0.0329}   &   0.9721$\pm${0.0294}     \\ \hline
			\multicolumn{1}{|l|}{Twitter}          &  0.9994$\pm${0.0038}  &  0.9995$\pm${0.0031}   &   0.9995$\pm${0.0031}     \\ \hline
			\multicolumn{1}{|l|}{Google}           &  0.9955$\pm${0.0089}  &  0.9952$\pm${0.0088}   &   0.9955$\pm${0.0089}     \\ \hline
			\multicolumn{1}{|l|}{Slack}            &  0.9254$\pm${0.2111}  &  0.9254$\pm${0.2111}   &   0.9254$\pm${0.2111}     \\ \hline
		\end{tabular}
	\end{table}
	
	\begin{table}[!t]
		\caption{Classification Recall}
		\label{tab:measures:recall}
		\renewcommand{\arraystretch}{1.3}
		\centering
		\begin{tabular}{l|l|l|l|}
			\cline{2-4}
			& \multicolumn{3}{c|}{\bfseries Technique} \\ \hline
			\multicolumn{1}{|c|}{\bfseries Dataset}          & \multicolumn{1}{c|}{\itshape\bfseries C4.5} & \multicolumn{1}{c|}{\itshape\bfseries RIPPER}                      & \multicolumn{1}{c|}{\itshape\bfseries PART}  \\ \hline
			\multicolumn{1}{|l|}{GHTraffic}        &  0.9837$\pm${0.0170}   &  0.9836$\pm${0.0169}    &  0.9837$\pm${0.0169}       \\ \hline
			\multicolumn{1}{|l|}{Twitter}          &  0.9996$\pm${0.0022}   &  0.9997$\pm${0.0019}    &  0.9997$\pm${0.0019}       \\ \hline
			\multicolumn{1}{|l|}{Google}           &  0.9972$\pm${0.0046}   &  0.9971$\pm${0.0045}    &  0.9972$\pm${0.0046}       \\ \hline
			\multicolumn{1}{|l|}{Slack}            &  0.9544$\pm${0.1290}   &  0.9544$\pm${0.1290}    &  0.9544$\pm${0.1290}       \\ \hline
		\end{tabular}
	\end{table}
	
	\begin{table}[!t]
		\caption{Tree Size or Number of Rules}
		\label{tab:measures:size}
		\renewcommand{\arraystretch}{1.3}
		\centering
		\begin{tabular}{l|l|l|l|}
			\cline{2-4}
			& \multicolumn{3}{c|}{\bfseries Technique} \\ \hline
			\multicolumn{1}{|c|}{\bfseries Dataset}          & \multicolumn{1}{c|}{\itshape\bfseries C4.5} & \multicolumn{1}{c|}{\itshape\bfseries RIPPER}                      & \multicolumn{1}{c|}{\itshape\bfseries PART}  \\ \hline
			\multicolumn{1}{|l|}{GHTraffic}        &  4.6735$\pm${9.4370}     &     2.1020$\pm${2.5269}     &   15.2041$\pm${11.5560}    \\ \hline
			\multicolumn{1}{|l|}{Twitter}          &  7.9538$\pm${0.6715}     &     3.0000$\pm${0.1768}     &   3.9692$\pm${0.3046}      \\ \hline
			\multicolumn{1}{|l|}{Google}           &  6.4118$\pm${0.9393}     &     2.8824$\pm${0.6966}     &   3.4706$\pm${1.1789}      \\ \hline
			\multicolumn{1}{|l|}{Slack}            &  10.0000$\pm${4.4401}    &     3.5000$\pm${0.9258}     &   3.6250$\pm${1.0607}      \\ \hline
		\end{tabular}
	\end{table}
	
	\medskip 
	According to descriptive statistics, all classifiers performed with an average accuracy of around 0.9541-0.9993, means that the error rate is low and most results are reliable. The observed averages of precision and recall are often quite close to the accuracy, so we can be reasonably confident that the classifiers are returning accurate results (high precision relates to a low false-positive rate) and most results are positive (high recall relates to a low false-negative rate). There is no significant difference in the measures of C4.5, RIPPER, and PART on each dataset. Especially, in Slack, each algorithm provides equivalent means for accuracy, precision, and recall. Further, the standard deviations in all those measures are quite low (range from 0.0022- 0.2111), confirms that there is low variation in the measurements for different training and testing sets in cross-validation. The average size of a tree or the number of rules produced by classifiers is around 2.1020-15.2041, means that the models are most compact and in a format that can easily be interpreted. 
	
	\medskip 
	While the classification algorithms obtained high measures for most targets, some obtained relatively low results. This is mainly due to insufficient training data (limited number of transaction sequences to reflect various behavioural patterns). The following is a detailed analysis of results along with a discussion of a few other circumstances which significantly affect the quality of predictions.
	
\begin{figure}[!t]
		\centerline{\includegraphics[width=\columnwidth]{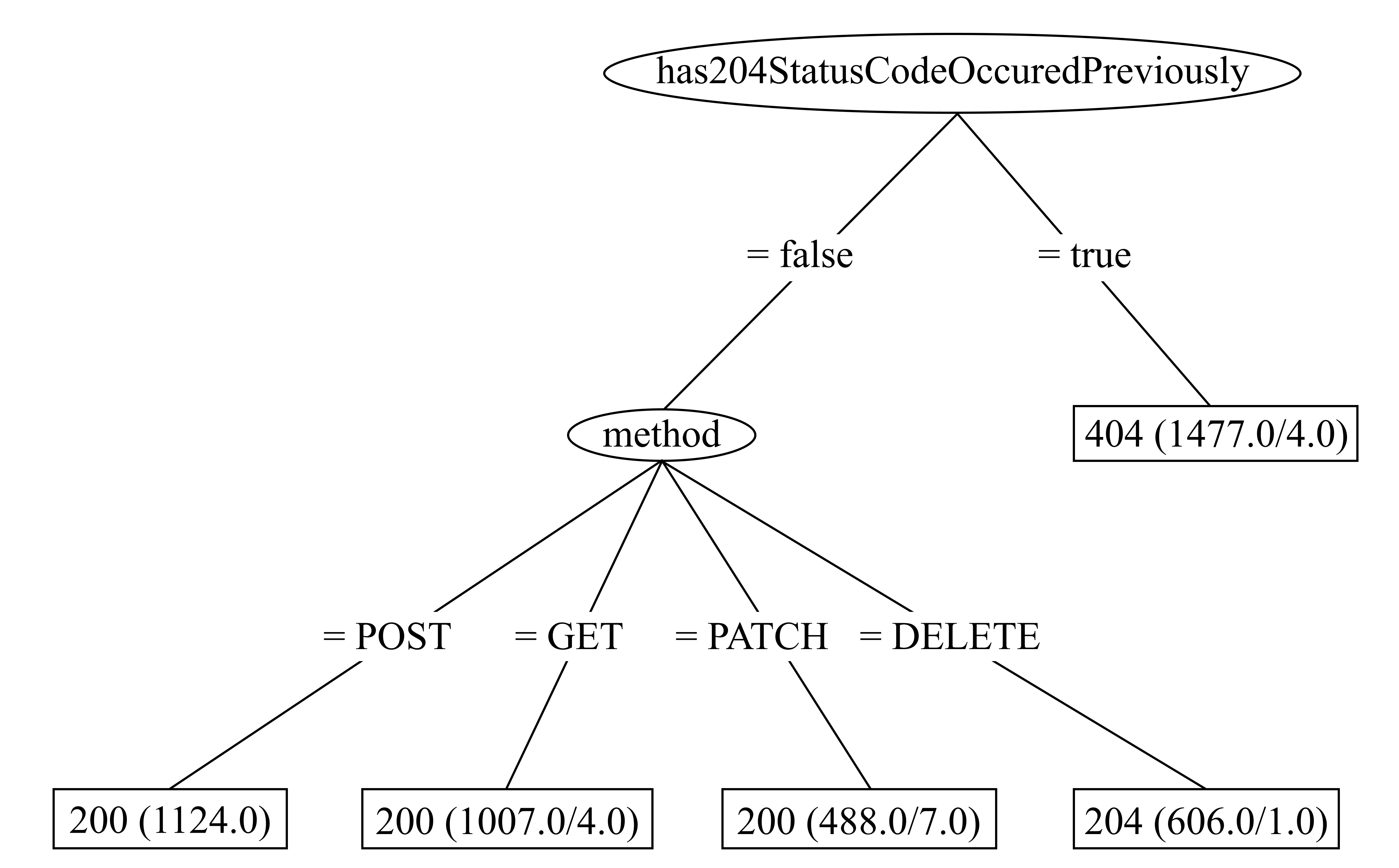}}
		\caption{The decision tree for Google Tasklists' status code on C4.5. Numbers in brackets indicate the total number of instances that fall into the particular leaf and the number of misclassified instances.}
		\label{fig:j48}
	\end{figure}

\begin{figure}[!t]
		\centerline{\includegraphics[width=\columnwidth]{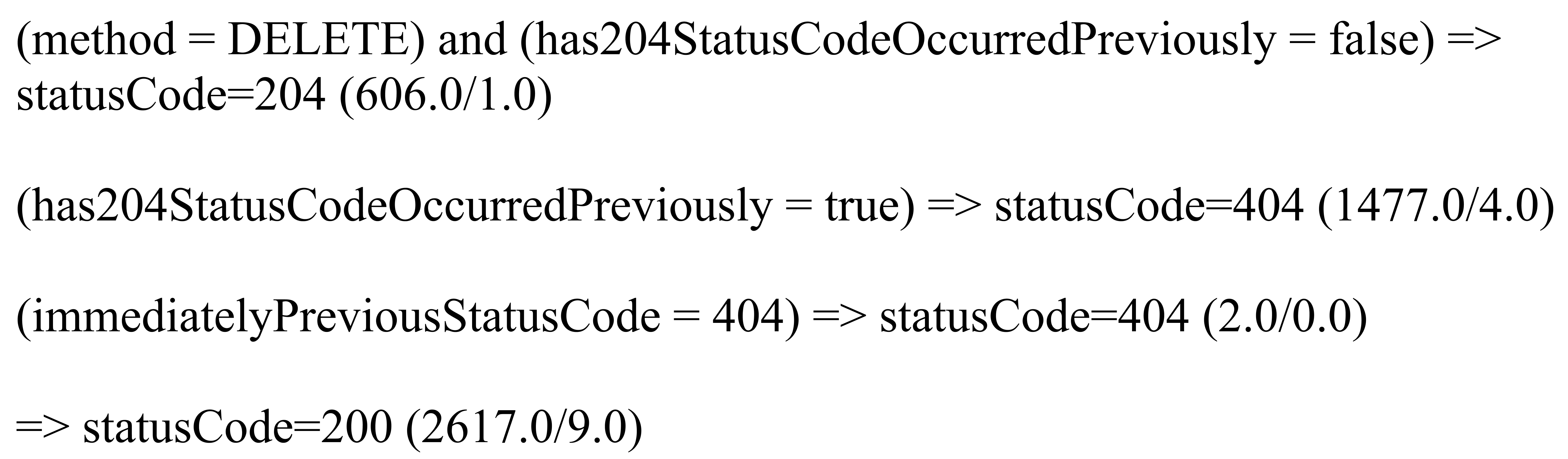}}
		\caption{The decision list for Google Tasklists' status code on RIPPER. Numbers in brackets indicate the total number of instances that classified into the particular rule and the number of misclassified instances.}
		\label{fig:jrip}
	\end{figure}

\begin{figure}[!t]
		\centerline{\includegraphics[width=0.81\columnwidth]{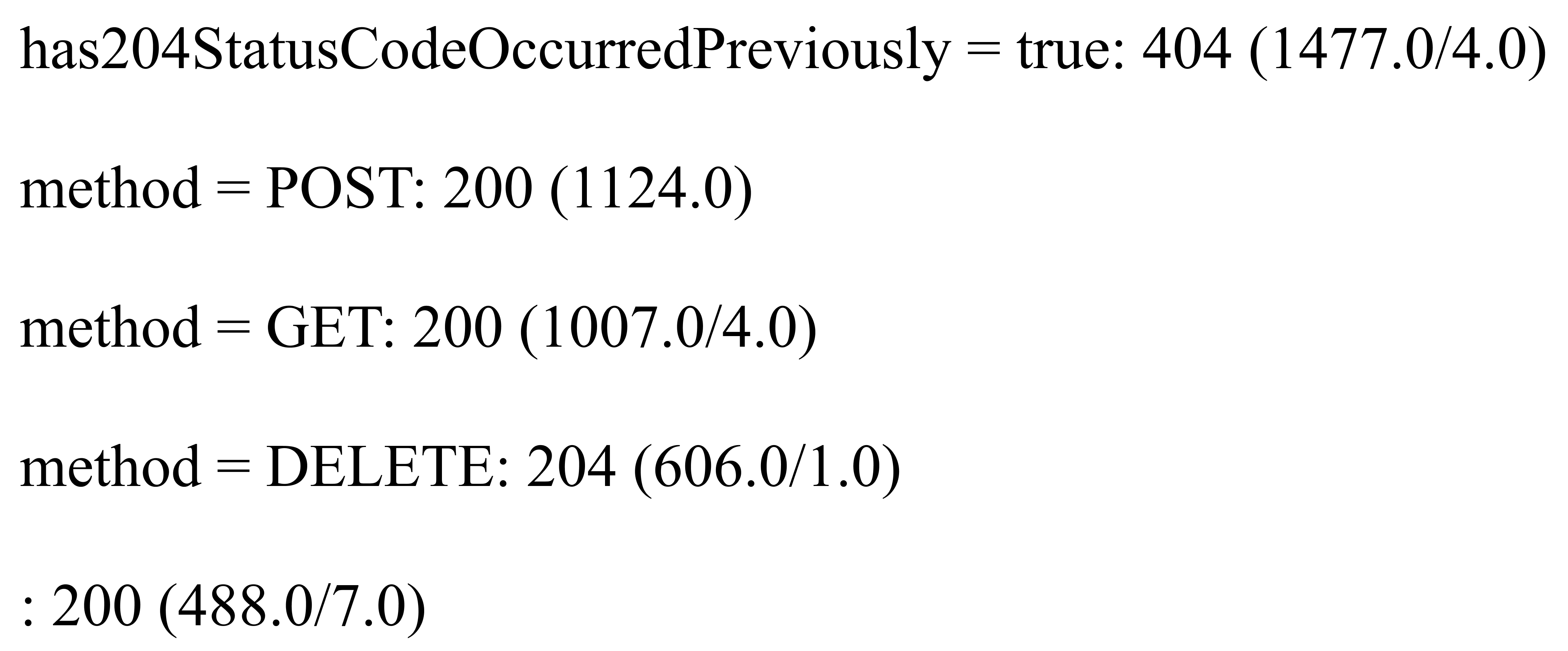}}
		\caption{The decision list for Google Tasklists' status code on PART. Numbers in brackets indicate the total number of instances that classified into the particular rule and the number of misclassified instances.}
		\label{fig:part}
	\end{figure}

	\medskip 
	Usually, HTTP services are supposed to embed success or failure of the request into the status code. Yet, some services always return a \textit{200} code (which indicates that the request made a successful call), even when it has unexpected behaviour, and include more substantive information about the status in the response body. For example, Slack implicitly returns \textit{200} but the response contains a boolean property \textit{ok}, indicating the status of the request. We noticed that the classifiers obtained high-performance results for all features associated with the state of the response. For example, C4.5, RIPPER, and PART models reached accuracy of 0.9996 for the \textit{responsejson:ok} target in Slack whereas the precision and recall rates are 1, while C4.5 built a tree with 8 leaves and each rule learner built a ruleset with 4 rules. Fig.~\ref{fig:j48} presents another sample classification model for \textit{statusCode} in Google Tasklists using C4.5. The constructed tree is size 7 and contains 5 leaves. Fig.~\ref{fig:jrip} and \ref{fig:part} show the models from RIPPER and PART where each ruleset contains 4 and 5 rules. All algorithms performed an accuracy of 0.9965 including the precision and recall rates are 0.9940 and 0.9970 respectively. 
	
	\medskip 
	Some of the targets relating to response body holds a fairly large set of distinct values (e.g., \textit{responsejson:message.text} in Slack), therefore ignored from learning. For most others, the classifiers are highly accurate and less sophisticated. But, for GHTraffic dataset, there are several models with 1 tree node or only the default class in the ruleset (it is because the standard pruning options are preventing the models from growing to restrain from overfitting) but with better performance scores (it is because the dataset contains a large number of instances with one value of the target and relatively few instances spread out over the rest of the values). For example, \textit{responsejson:closed{\_}by.id} contains 99.98\% of instances with \textit{not-exist} value, so the classifier could predict every \textit{responsejson:closed{\_}by.id} as \textit{no-exist} and achieve a 0.9998 accuracy. Even if the target has a few distinct values, this may not lead to reliable classification models due to the high impact of the majority value. The GHTraffic payload includes several features with such imbalanced value distributions which not most appropriate to be predicted. We further observed that different targets relating to the same feature (usually have the same value distributions) generate the same classification model. For example, users on Twitter are uniquely identified by properties such as \textit{id}, \textit{id{\_}str}, \textit{name}, etc., and the associated target attributes \textit{responsejson:user.id}, \textit{responsejson:user.id{\_}str}, etc., formed identical classifiers with the same measurements. 
	
	\medskip 
	Vast majority of target attributes relating to response headers in all four datasets are either single-valued or with a large number of distinct values, thus ignored from building classifiers. All other built models are accurate and easier to interpret in most datasets, i.e. C4.5, RIPPER, and PART provide the accuracy, precision, and recall rate of 1 for \textit{responseheader:Cache-Control} in GHTraffic while C4.5 output a tree with 9 leaves and size 12, RIPPER and PART built a ruleset with 5 and 9 rules. Except, for Slack with all algorithms, there is a significant variation in performance values in \textit{responseheader:X-slack-router} (which is the only target remaining for prediction). The reached accuracy, precision, and recall are 0.6351, 0.4030, 0.6350, respectively (very low compared to other results). C4.5 output a tree with size 1. For rule algorithms, RIPPER produced a classifier with 2 rules and PART with 1 (which is the default class). This is due to there being imbalanced value distribution. Also, this results in relatively low average scores for Slack with a notable value distribution.
	
	\medskip 
	Overall, the achieved results reveal that the significance of SML algorithms in training accurate, human-readable models for predicting the key features of HTTP service responses (e.g. status, response headers, response body). It also confers the usefulness of the proposed attributes in building classification models.
	
	\section{Threats To Validity} \label{sec:validity}
	
	As described in Section~\ref{sec:datasets}, most of the datasets used in the study were collected from recording the network traffic through fuzzing REST APIs. This leaves the possibility that datasets do not reflect realistic workloads, thereby, research results might be not realistic and cannot be generalised. However, we extracted datasets from the most successful active Web services and synthesised using a well-defined process. All API interactions were derived according to syntax and semantics of HTTP and the underline service, also implemented a wide set of HTTP features. Each dataset was large enough to facilitate the research described. We were able to preserve the behaviour that exists in the real HTTP traffic as much as possible and believed that all datasets reflect the state-of-the-art use of HTTP-based services in general. Therefore, we are certain that research outcomes were accurate and could be applied to other HTTP services that were not studied. 
	
	\medskip 
	The models have only been evaluated using metrics such as accuracy, precision, and recall (to assess the correctness), and model size (to assess the comprehensibility). Those might not be sufficient to assess how suitable the SML algorithms will be in producing mock response skeletons. Further analysis is expected in the future (i.e. evaluate models by a targeted group of end-users to determine the extent of its usability). 
	
	\section{Conclusion} \label{sec:conclusion}
	
	In this paper, we have studied the potential of symbolic machine learning algorithms in producing mock response skeletons of HTTP-based services. The chosen algorithms demonstrate the suitability of producing accurate semantic models. Part of our motivation was to produce  skeletons for mocked services to be customised by engineers. The usability of the generated models has not yet been assessed, this needs to be addressed in a future empirical study with end-users. 
	
	\begingroup\let\itshape\upshape
	\bibliographystyle{IEEEtran}
	\def\IEEEbibitemsep{0pt plus .5pt}
	\bibliography{IEEEabrv,bibliography}

\begin{thebibliography}{10}
\providecommand{\url}[1]{#1}
\csname url@samestyle\endcsname
\providecommand{\newblock}{\relax}
\providecommand{\bibinfo}[2]{#2}
\providecommand{\BIBentrySTDinterwordspacing}{\spaceskip=0pt\relax}
\providecommand{\BIBentryALTinterwordstretchfactor}{4}
\providecommand{\BIBentryALTinterwordspacing}{\spaceskip=\fontdimen2\font plus
\BIBentryALTinterwordstretchfactor\fontdimen3\font minus
  \fontdimen4\font\relax}
\providecommand{\BIBforeignlanguage}[2]{{%
\expandafter\ifx\csname l@#1\endcsname\relax
\typeout{** WARNING: IEEEtran.bst: No hyphenation pattern has been}%
\typeout{** loaded for the language `#1'. Using the pattern for}%
\typeout{** the default language instead.}%
\else
\language=\csname l@#1\endcsname
\fi
#2}}
\providecommand{\BIBdecl}{\relax}
\BIBdecl

\bibitem{papazoglou2003service}
M.~P. Papazoglou, ``Service-oriented computing: Concepts, characteristics and
  directions,'' in \emph{Proc. WISE'03}.\hskip 1em plus 0.5em minus 0.4em\relax
  IEEE, 2003.

\bibitem{christensen2001web}
\BIBentryALTinterwordspacing
E.~Christensen \emph{et~al.}, ``{Web services description language (WSDL)
  1.1},'' 2001, accessed Jul. 2 2019. [Online]. Available:
  \url{https://w3.org/TR/wsdl}
\BIBentrySTDinterwordspacing

\bibitem{box1999soap}
\BIBentryALTinterwordspacing
D.~Box \emph{et~al.}, ``{Simple object access protocol (SOAP) 1.1},'' 2000,
  accessed Jul. 2 2019. [Online]. Available:
  \url{https://w3.org/TR/2000/NOTE-SOAP-20000508/}
\BIBentrySTDinterwordspacing

\bibitem{fielding2000architectural}
R.~T. Fielding and R.~N. Taylor, \emph{Architectural styles and the design of
  network-based software architectures}.\hskip 1em plus 0.5em minus 0.4em\relax
  University of California, Irvine Irvine, USA, 2000, vol.~7.

\bibitem{freeman2004mock}
S.~Freeman, T.~Mackinnon, N.~Pryce, and J.~Walnes, ``Mock roles, not objects,''
  in \emph{Proc. OOPSLA'04}.\hskip 1em plus 0.5em minus 0.4em\relax ACM, 2004.

\bibitem{michelsen2012service}
J.~Michelsen and J.~English, ``What is service virtualization?'' in
  \emph{Service Virtualization}.\hskip 1em plus 0.5em minus 0.4em\relax
  Springer, 2012.

\bibitem{michalski2013machine}
R.~S. Michalski, J.~G. Carbonell, and T.~M. Mitchell, \emph{Machine learning:
  An artificial intelligence approach}.\hskip 1em plus 0.5em minus 0.4em\relax
  Springer Science \& Business Media, 2013.

\bibitem{holzinger2018machine}
A.~Holzinger, ``From machine learning to explainable ai,'' in \emph{Proc.
  DISA'18}.\hskip 1em plus 0.5em minus 0.4em\relax IEEE, 2018.

\bibitem{hall2009weka}
M.~Hall, E.~Frank, G.~Holmes, B.~Pfahringer, P.~Reutemann, and I.~H. Witten,
  ``The weka data mining software: an update,'' \emph{ACM SIGKDD explorations
  newsletter}, vol.~11, no.~1, 2009.

\bibitem{parasoft}
\BIBentryALTinterwordspacing
``{Parasoft Virtualize},'' accessed Jul. 2 2019. [Online]. Available:
  \url{https://parasoft.com/products/virtualize}
\BIBentrySTDinterwordspacing

\bibitem{ca}
\BIBentryALTinterwordspacing
``{CA Service Virtualization},'' accessed Jul. 2 2019. [Online]. Available:
  \url{https://ca.com/us/products/ca-service-virtualization.html}
\BIBentrySTDinterwordspacing

\bibitem{wiremock}
\BIBentryALTinterwordspacing
``{Wiremock},'' accessed Jul. 2 2019. [Online]. Available:
  \url{http://wiremock.org/}
\BIBentrySTDinterwordspacing

\bibitem{hoverfly}
\BIBentryALTinterwordspacing
``{Hoverfly},'' accessed Jul. 2 2019. [Online]. Available:
  \url{https://hoverfly.io/}
\BIBentrySTDinterwordspacing

\bibitem{du2016opaque}
M.~Du, ``Opaque response generation enabling automatic creation of virtual
  services for service virtualisation,'' \emph{arXiv preprint
  arXiv:1608.04885}, 2016.

\bibitem{versteeg2017entropy}
S.~C. Versteeg, J.~S. Bird, N.~A. Hastings, M.~Du, and J.-D. Dahan, ``Entropy
  weighted message matching for opaque service virtualization,'' 2017, uS
  Patent 9,582,399.

\bibitem{needleman1970general}
S.~B. Needleman and C.~D. Wunsch, ``A general method applicable to the search
  for similarities in the amino acid sequence of two proteins,'' \emph{Journal
  of molecular biology}, vol.~48, no.~3, 1970.

\bibitem{eniser2018fancymock}
H.~F. Eniser, A.~Sen, and S.~O. Polat, ``Fancymock: creating virtual services
  from transactions,'' in \emph{Proc. SAC'18}.\hskip 1em plus 0.5em minus
  0.4em\relax ACM, 2018.

\bibitem{enicser2018testing}
H.~F. Eni{\c{s}}er and A.~Sen, ``Testing service oriented architectures using
  stateful service visualization via machine learning,'' in \emph{Proc.
  AST'18}.\hskip 1em plus 0.5em minus 0.4em\relax ACM, 2018.

\bibitem{bozkurt2010testing}
M.~Bozkurt, M.~Harman, Y.~Hassoun \emph{et~al.}, ``Testing web services: A
  survey,'' \emph{Department of Computer Science, King’s College London,
  Tech. Rep. TR-10-01}, 2010.

\bibitem{kumar2015empirical}
A.~Kumar and M.~Singh, ``An empirical study on testing of soa based services,''
  \emph{International Journal of Information Technology and Computer Science},
  vol.~7, no.~1, 2015.

\bibitem{canfora2007service}
G.~Canfora and M.~Di~Penta, ``Service-oriented architectures testing: A
  survey,'' in \emph{Software Engineering}.\hskip 1em plus 0.5em minus
  0.4em\relax Springer, 2007, pp. 78--105.

\bibitem{arcuri2018evomaster}
A.~Arcuri, ``Evomaster: Evolutionary multi-context automated system test
  generation,'' in \emph{Proc. ICST'18}.\hskip 1em plus 0.5em minus 0.4em\relax
  IEEE, 2018.

\bibitem{lamela2013towards}
P.~Lamela~Seijas, H.~Li, and S.~Thompson, ``Towards property-based testing of
  restful web services,'' 2013.

\bibitem{singhal2014generation}
A.~Singhal, A.~Bansal \emph{et~al.}, ``Generation of test oracles using neural
  network and decision tree model,'' in \emph{Proc. Confluence'14}.\hskip 1em
  plus 0.5em minus 0.4em\relax IEEE, 2014.

\bibitem{briand2008using}
L.~C. Briand, Y.~Labiche, and Z.~Bawar, ``Using machine learning to refine
  black-box test specifications and test suites,'' in \emph{Proc.
  QSIC'08}.\hskip 1em plus 0.5em minus 0.4em\relax IEEE, 2008.

\bibitem{kanewala2016predicting}
U.~Kanewala, J.~M. Bieman, and A.~Ben-Hur, ``Predicting metamorphic relations
  for testing scientific software: a machine learning approach using graph
  kernels,'' \emph{Software testing, verification and reliability}, vol.~26,
  no.~3, 2016.

\bibitem{hewson2015performance}
F.~Hewson, J.~Dietrich, and S.~Marsland, ``Performance regression testing on
  the java virtual machine using statistical test oracles,'' in \emph{Proc.
  ASWEC'15}.\hskip 1em plus 0.5em minus 0.4em\relax IEEE, 2015.

\bibitem{hoschele2016mining}
M.~H{\"o}schele and A.~Zeller, ``Mining input grammars from dynamic taints,''
  in \emph{Proc. ASE'16}.\hskip 1em plus 0.5em minus 0.4em\relax ACM, 2016.

\bibitem{atlidakis2019restler}
V.~Atlidakis, P.~Godefroid, and M.~Polishchuk, ``Restler: Stateful rest api
  fuzzing,'' in \emph{Proc. ICSE'19}.\hskip 1em plus 0.5em minus 0.4em\relax
  IEEE Press, 2019.

\bibitem{bhagya2018ghtraffic}
T.~Bhagya, J.~Dietrich, H.~Guesgen, and S.~Versteeg, ``Ghtraffic: A dataset for
  reproducible research in service-oriented computing,'' in \emph{Proc.
  ICWS'18}.\hskip 1em plus 0.5em minus 0.4em\relax IEEE, 2018.

\bibitem{galiegue2013json}
F.~Galiegue and K.~Zyp, ``{JSON Schema: core definitions and terminology
  draft-zyp-json-schema-04},'' \emph{Working Draft}, 2013.

\bibitem{quinlan1986induction}
J.~R. Quinlan, ``Induction of decision trees,'' \emph{Machine learning},
  vol.~1, no.~1, 1986.

\bibitem{quinlan2014c4}
J.~Quinlan, \emph{C4. 5: programs for machine learning}.\hskip 1em plus 0.5em
  minus 0.4em\relax Elsevier, 2014.

\bibitem{breiman1984classification}
L.~Breiman, J.~Friedman, R.~Olshen, and C.~Stone, ``Classification and
  regression trees. wadsworth int,'' \emph{Group}, vol.~37, no.~15, 1984.

\bibitem{freund1999alternating}
Y.~Freund and L.~Mason, ``The alternating decision tree learning algorithm.''

\bibitem{ho1995random}
T.~K. Ho, ``Random decision forests,'' in \emph{Proc. ICDAR'95}.\hskip 1em plus
  0.5em minus 0.4em\relax IEEE, 1995.

\bibitem{frank1998generating}
E.~Frank and I.~H. Witten, ``Generating accurate rule sets without global
  optimization,'' 1998.

\bibitem{cohen1995fast}
W.~W. Cohen, ``Fast effective rule induction,'' in \emph{Machine Learning Proc.
  1995}.\hskip 1em plus 0.5em minus 0.4em\relax Elsevier, 1995.

\bibitem{holte1993very}
R.~C. Holte, ``Very simple classification rules perform well on most commonly
  used datasets,'' \emph{Machine learning}, vol.~11, no.~1, 1993.

\bibitem{Kohavi1995}
R.~Kohavi, ``The power of decision tables,'' in \emph{Proc. ECML'95}.\hskip 1em
  plus 0.5em minus 0.4em\relax Springer, 1995.

\bibitem{clark1991rule}
P.~Clark and R.~Boswell, ``Rule induction with cn2: Some recent improvements,''
  in \emph{Proc. EWSL'91}.\hskip 1em plus 0.5em minus 0.4em\relax Springer,
  1991.

\bibitem{kotsiantis2007supervised}
S.~B. Kotsiantis, I.~Zaharakis, and P.~Pintelas, ``Supervised machine learning:
  A review of classification techniques,'' \emph{Emerging artificial
  intelligence applications in computer engineering}, 2007.

\bibitem{witten2016data}
I.~H. Witten, E.~Frank, M.~A. Hall, and C.~J. Pal, \emph{Data Mining: Practical
  machine learning tools and techniques}.\hskip 1em plus 0.5em minus
  0.4em\relax Morgan Kaufmann, 2016.

\bibitem{berners1998uniform}
T.~Berners-Lee, R.~Fielding, L.~Masinter \emph{et~al.}, ``Uniform resource
  identifiers (uri): Generic syntax,'' 1998.

\end{thebibliography}
	\endgroup

\end{document}